\documentclass[twocolumn]{aastex6}
\bibstyle{aasjournal}

\turnoffedit

\newcommand{\xmm}{{\it XMM-Newton}}
\newcommand{\xmms}{{\it XMM}}
\newcommand{\chandra}{{\it Chandra}}

\newcommand{\rosat}{{\it ROSAT}}
\newcommand{\acis}{{\rm ACIS}}
\newcommand{\xspec}{\rm XSPEC}

\newcommand{\HII}{\mbox{H{\sc ii}}}
\newcommand{\nhunit}{$\mbox{cm}^{-2}$}
\newcommand{\nobs}{33}

\newcommand{\sna}{SN 1987A}

\shorttitle{Chandra Observations of \sna}
\shortauthors{Frank et al.}

\slugcomment{{\sc Accepted to ApJ:} August 6, 2016}

\begin{document}
\title{Chandra Observes the End of an Era in \sna}

\author{Kari A. Frank}
\affil{Department of Astronomy and Astrophysics, Pennsylvania State University, University Park, PA 16802, USA}
\email{kafrank@psu.edu}
\author{Svetozar A. Zhekov}
\affil{Institute of Astronomy and National Astronomical Observatory, 72 Tsarigradsko Chaussee Blvd., Sofia 1784, Bulgaria}
\author{Sangwook Park}
\affil{Department of Physics, University of Texas at Arlington, Arlington, TX 76019, USA}
\author{Richard McCray}
\affil{Department of Astronomy, University of California, Berkeley, CA 94720-3411, USA}
\author{Eli Dwek}
\affil{Observational Cosmology Laboratory, Code 665, NASA Goddard Space Flight Center, Greenbelt, MD 20771, USA}
\and
\author{David N. Burrows}
\affil{Department of Astronomy and Astrophysics, Pennsylvania State University, University Park, PA 16802, USA}

\begin{abstract}
Updated imaging and photometric results from \chandra\ observations of \sna, covering the last 16 years, are presented.  We find that the 0.5-2 keV light curve has remained constant at $\sim$$8\times10^{-12}$ erg s$^{-1}$cm$^{-2}$  since 9500 days, with the 3-8 keV light curve continuing to increase until at least 10000 days.  The expansion rate of the ring is found to be energy dependent, such that after day 6000 the ring expands faster in the 2-10 keV band than it does at energies $<2$ keV.  Images show a reversal of the east-west asymmetry between 7000 and 8000 days after the explosion.  The latest images suggest the southeastern side of the equatorial ring is beginning to fade.  Consistent with the latest optical and infrared results, our \chandra\ analysis indicates the blast wave is now leaving the dense equatorial ring\edit1{, which marks the beginning of a major change in the evolutionary phase of the supernova remnant 1987A.}
\end{abstract}



\section{Introduction}
\label{section:intro}
As the only nearby supernova observed in the last 400 years, \sna\ provides the unique opportunity to study in detail the first decades of a supernova remnant's development. 
Over the last 28 years, \sna\ has been evolving on timescales of months to years; regular monitoring at multiple wavelengths has therefore been crucial for tracking these changes and understanding the development of the newborn remnant. Early optical observations revealed an unusual triple ring system \citep{Burrows1995} consisting of two outer rings and a bright equatorial ring (ER) that together form an hourglass structure. The inner ER is embedded in a larger, lower density \HII\ region \citep{Chevalier1995}. This circumstellar structure is likely the result of interaction between a slow, dense red supergiant wind and the faster, lighter blue supergiant wind \citep{Luo1991,Wang1992}.  The origin of this ring morphology is unclear, but may point toward either a binary-merger \edit1{\citep{Blondin1993,Morris2007,Morris2009}} or fast-rotator \edit1{\citep{Chita2008}} scenario for the progenitor star.

X-ray emission was detected by \rosat\ 1400 days after the supernova, as the blast wave first encountered the \HII\ region interior to the ER \citep{Hasinger1996}.  Starting about 4000 days after the supernova, the blast wave began interacting with very dense ($n_e\sim10^4$ cm$^{-3}$) clumps protruding from the inner edge of the ER, visible as `hot spots' in optical images \citep{Sugerman2002} with associated X-ray emission seen with \chandra\ \citep{Burrows2000,Park2002}.  Subsequent \chandra\ observations revealed impact with the main body of the ER around day 6000, resulting in a sudden decrease in the blast wave velocity \citep{Racusin2009} and a dramatic increase in the soft X-ray flux \citep{Park2004,Park2005} as the dense material of the clumps and the ring were shock-heated. The X-ray flux has continued to rise since day 6000, indicating continuous interaction with the ER \citep{Park2005,Park2011,Maggi2012,Helder2013}.

However, recent optical results suggest a transition is occurring.  \citet{Fransson2015} have noted the appearance of diffuse emission and hot spots outside of the ER, beginning in 2013 ($\sim$9500 days).  The optical hot spots, associated with the densest clumps in the ER, have also been fading since day $\sim$8000. \citet{Fransson2015} interpret this as the beginning of the clumps' destruction by the expanding supernova ejecta.  The mid-infrared emission from dust in the ER also began to fade around this time \citep{Arendt2016}.

The X-ray light curve is a straightforward and powerful probe of the circumstellar medium (CSM) density, especially given the frequent sampling with \chandra\ $-$ approximately every 6 months with \chandra\ \acis\ since day 4600.  These observations have been complemented by less frequent \xmm\ observations and earlier \rosat\ fluxes. The spatially resolved \chandra\ observations also provide measurements of the ring expansion velocity and allow comparison with the morphology at optical, infrared, and radio wavelengths.  Both the X-ray light curve and images have been key ingredients for constraining models of \sna\ evolution \citep{Chevalier1995,Borkowski1997b,Zhekov2010,Dewey2012,Orlando2015}.  These models predict a flattening of the light curve once the blast wave leaves the main body of the ER. The subsequent slope of the soft and hard X-ray light curves will depend on the density structure of the CSM beyond the ring \citep[e.g.][]{Park2011}, which is unknown, as well as the properties of the ejecta, which will likely begin to dominate the X-ray emission a few years after the blast wave leaves the ER \citep{Orlando2015}.

In this paper, we present updated \chandra\ light curves, as well as radial expansion measurements and images, covering up to 10433 days after the supernova (through 2015 September).  Details of the observations and data reduction are given in \S\ref{section:observationsreduction} with the analysis and results described in \S\ref{section:analysis}. A discussion of these results in light of recent multiwavelength and modeling studies is provided in \S\ref{section:discussion}, with a summary and conclusions in \S\ref{section:conclusions}.  We assumed a distance of 51.4 kpc to \sna\ \citep{Panagia1999}. \edit1{All uncertainties are 90\% confidence unless stated otherwise.}

\section{Observations and Data Reduction}
\label{section:observationsreduction}
The \nobs\ \chandra\ observations used in this work are described in Table \ref{table:observations}.  For observations through 2012, the observing configuration changed several times to reduce the effects of photon pileup, as described in \citet{Helder2013}.  There have been six new observations since 2012.   Obs ID 16757 utilized the HRC-S/LETG, which is not affected by the molecular contamination on the ACIS optical blocking filter.   The other five utilized the same ACIS/HETG configuration as previous observations, but with the observation moved back to the center of the chip to minimize the effects of contaminant absorption, which is lower and better characterized near the center of the detector. We obtained spectral and imaging results from all observations, with a few exceptions.  We did not analyze the spectra of the earliest two observations, Obs IDs 1387 and 122, as the focal plane temperature was $-110$ C and thus there is no correction for charge transfer inefficiency.  The day 7800 and 8000 epochs included observations both with and without the HETG.  For these epochs we used the 0th order images and spectra from HETG observations for spectral analysis, as the use of the grating reduces pileup, and we used the bare ACIS observation for the imaging analysis, as these contain more counts.  From Obs ID 16757 we did not use the HRC imaging data, only the 0.5-2 keV flux from the dispersed LETG spectrum.    

Data reduction followed essentially the same procedure used in previous works \citep{Burrows2000,Racusin2009,Park2011,Helder2013}. ACIS 0th order spectra were extracted from a 4\farcs38 radius circular region, with corresponding background spectra extracted from a concentric annulus with inner radius of 6\farcs2 and outer radius of 12\farcs4, using CIAO 4.7 and CALDB 4.6.5.  While the instrument configuration was chosen to minimize pileup, \citet{Helder2013} found that modest pileup was still present in many of the observations.  We followed the same procedure used in \citet{Helder2013} to correct the 0th order spectra for this pileup.  The dispersed LETG spectra (positive and negative orders) of Obs ID 16757 were extracted and binned with a minimum of 30 counts per bin. 

\xmms\ EPIC-pn fluxes through 2011 are available in \citet{Haberl2006} and \citet{Maggi2012}.  However, for consistency with our \chandra\ analysis and to obtain fluxes for the \edit1{unpublished} 2012 and 2014 \xmms\ observations, we reanalyzed  spectra from all EPIC-pn observations since 2001 (day 5100).  Details of the observations are shown in Table \ref{table:observations}. EPIC-pn spectra were extracted using SAS version 14, but otherwise following the procedure used by \citet{Maggi2012}.  It was also necessary to correct most of the EPIC-pn spectra for pileup, using the method described in the relevant SAS data analysis thread\footnote{\href{http://www.cosmos.esa.int/web/xmm-newton/sas-thread-epatplot}{\url http://www.cosmos.esa.int/web/xmm-newton/sas-thread-epatplot}}.  

Images were obtained using the same approach as our previous works \citep{Burrows2000,Racusin2009,Park2011,Helder2013}.  Pixel randomization added by the \chandra\ software to older data sets was removed. We used split-pixel events to achieve sub-pixel resolution \citep{Mori2001}; a critical step as \sna\ is barely resolved in raw ACIS images with a diameter of 1\farcs5.  Images were then deconvolved with the point spread function of the telescope using the Lucy-Richardson iterative deconvolution algorithm \citep{Lucy1974,Richardson1972} and smoothed by convolving with a gaussian (FWHM$\sim$0\farcs1).  Use of the HETG changes the spectral response compared to that of the bare ACIS detector, biasing the 0th order images toward somewhat higher energies.  We have two epochs which include observations both with and without the HETG. Comparing the imaging results for these observations, we found that the differences are not significant enough to change our overall results or conclusions.
\newpage
\newcommand\tfn{\,\tablenotemark{$^{\mbox{a}}$}}
\onecolumngrid
\begin{deluxetable}{cccccccc}~
\tabletypesize{\footnotesize}
\tablecolumns{8}
\tablecaption{\chandra\ and \xmms\ Observations, \label{table:observations} Fluxes, Radii}
\tablewidth{0pt}
\tablehead{\colhead{Obs ID} & \colhead{Date} & \colhead{Age} & \colhead{Instrument} & \colhead{Exposure} & \colhead{0.5-2.0 keV Flux} & \colhead{3.0-8.0 keV Flux } &\colhead{Radius} \\ 
 &  & (days) & & (ks) & (10$^{-13}$erg s$^{-1}$cm$^{-2}$) & (10$^{-13}$erg s$^{-1}$cm$^{-2}$) &(arcsec)} \\ 
\startdata
1387 & 1999-10-06 & 4608 & ACIS/HETG & 68.9 & \ldots & \ldots & 0.586$\pm$0.033\\ 
122 & 2000-01-17 & 4711 & ACIS & 8.6 & \ldots & \ldots & 0.597$\pm$0.026\\ 
1967 & 2000-12-07 & 5036 & ACIS & 98.8 & 2.71$_{-0.05}^{+0.05}$ & 0.72$_{-0.09}^{+0.10}$ & 0.664$\pm$0.007 \\ 
1044 & 2001-04-25 & 5175 & ACIS & 17.8 & 3.00$_{-0.15}^{+0.15}$ & 0.91$_{-0.23}^{+0.29}$ & 0.681$\pm$0.019\\
2831 & 2001-12-12 & 5406 & ACIS & 49.4 & 4.11$_{-0.11}^{+0.11}$ & 0.91$_{-0.21}^{+0.21}$ & 0.693$\pm$0.008\\
2832 & 2002-05-15 & 5560 & ACIS & 44.3 & 5.09$_{-0.13}^{+0.13}$ & 1.05$_{-0.21}^{+0.21}$ & 0.698$\pm$0.008\\
3829 & 2002-12-31 & 5789 & ACIS & 49.0 & 7.18$_{-0.14}^{+0.15}$ & 1.21$_{-0.19}^{+0.20}$ & 0.718$\pm$0.007\\
3830 & 2003-07-08 & 5978 & ACIS & 45.3 & 9.20$_{-0.20}^{+0.20}$ & 1.52$_{-0.27}^{+0.27}$ & 0.728$\pm$0.007\\
4614 & 2004-01-02 & 6157 & ACIS & 46.5 & 11.26$_{-0.23}^{+0.23}$ & 1.73$_{-0.29}^{+0.29}$& 0.743$\pm$0.006\\
4615 & 2004-07-22 & 6359 & ACIS & 48.8 &  14.45$_{-0.21}^{+0.20}$ & 1.80$_{-0.22}^{+0.24}$ & 0.743$\pm$0.005\\
5579 & 2005-01-09 & 6530 & ACIS & 31.9 & 17.28$_{-0.26}^{+0.27}$ & 1.87$_{-0.24}^{+0.24}$ & 0.745$\pm$0.005\\
5580 & 2005-07-11 & 6713 & ACIS & 23.8 & 21.4$_{-0.35}^{+0.36}$ & 2.34$_{-0.30}^{+0.36}$ & 0.753$\pm$0.006\\
6668 & 2006-01-28 & 6913 & ACIS & 42.3 & 26.3$_{-0.28}^{+0.28}$ & 2.76$_{-0.23}^{+0.25}$& 0.750$\pm$0.004\\
6669 & 2006-07-27 & 7094 & ACIS & 42.3 & 31.25$_{-0.35}^{+0.37}$ & 3.13$_{-0.29}^{+0.29}$& 0.757$\pm$0.004\\
7636 & 2007-01-19 & 7270 & ACIS & 33.5 & 37.41$_{-0.38}^{+0.40}$ & 3.59$_{-0.32}^{+0.30}$& 0.764$\pm$0.004\\
7637 & 2007-07-13 & 7445 & ACIS & 25.7 & 42.22$_{-0.50}^{+0.51}$ & 3.24$_{-0.36}^{+0.44}$ & 0.769$\pm$0.004\\
9142 & 2008-01-09 & 7624 & ACIS & 6.6 & 46.41$_{-0.96}^{+0.96}$ & 3.45$_{-0.66}^{+0.82}$& 0.772$\pm$0.008\\
9144 & 2008-07-01 & 7799 & ACIS/HETG & 42.0 & 50.18$_{-1.52}^{+1.53}$ & 4.94$_{-0.72}^{+0.61}$ & \ldots\\
9143 & 2008-07-04 & 7802 & ACIS & 8.6 & \ldots & \ldots & 0.766$\pm$0.006\\
10130 & 2009-01-05 & 7986 & ACIS & 6.0 & \ldots & \ldots & 0.786$\pm$0.008\\
10855 & 2009-01-18 & 8000 & ACIS/HETG & 18.8 & 55.58$_{-2.49}^{+2.47}$ & 4.19$_{-0.97}^{+1.03}$ & \ldots\\
10222 & 2009-07-06 & 8169 & ACIS/HETG & 24.4 & 61.36$_{-2.27}^{+2.32}$ & 5.17$_{-0.79}^{+1.04}$& 0.787$\pm$0.012\\
11090\tfn & 2010-03-28 & 8433 & ACIS/HETG & 24.6 & 63.83$_{-2.34}^{+2.39}$ & 5.61$_{-0.84}^{+0.77}$ & 0.788$\pm$0.012\\ 
13131\tfn & 2010-09-28 & 8617 & ACIS/HETG & 26.5 & 68.63$_{-2.35}^{+2.36}$ & 5.75$_{-0.88}^{+0.76}$ & 0.781$\pm$0.012\\ 
12539\tfn & 2011-03-25 & 8796 & ACIS/HETG & 52.2 & 71.25$_{-1.91}^{+1.90}$ & 7.29$_{-0.61}^{+0.61}$ & 0.787$\pm$0.008\\ 
12540\tfn & 2011-09-21 & 8975 & ACIS/HETG & 37.5 & 74.92$_{-2.16}^{+2.19}$ & 7.45$_{-0.74}^{+0.73}$ & 0.802$\pm$0.010\\ 
13735\tfn & 2012-03-28 & 9165 & ACIS/HETG & 42.9 & 80.71$_{-2.20}^{+2.07}$ & 8.91$_{-0.76}^{+0.77}$ & 0.799$\pm$0.009\\ 
14697 & 2013-03-21 & 9523 & ACIS/HETG & 67.6 & 77.48$_{-1.66}^{+1.68}$ & 9.71$_{-0.64}^{+0.73}$ & 0.810$\pm$0.007\\
14698 & 2013-09-28 & 9713 & ACIS/HETG & 68.5 & 81.00$_{-1.55}^{+1.55}$ & 11.17$_{-0.67}^{+0.67}$& 0.814$\pm$0.007 \\
15809 & 2014-03-19 & 9885 & ACIS/HETG & 70.5 & 78.96$_{-1.73}^{+1.82}$ & 11.53$_{-0.74}^{+0.71}$ & 0.813$\pm$0.007\\
15810 & 2014-09-20 & 10071 & ACIS/HETG & 47.9 & 80.56$_{-2.32}^{+2.33}$ & 11.50$_{-0.96}^{+0.92}$& 0.830$\pm$0.009\\
16757 & 2015-03-14 & 10246 & HRC/LETG & 67.6 & 81.21$_{-2.63}^{+2.68}$  & \ldots & \ldots\\
16756 & 2015-09-17 & 10433 & ACIS/HETG & 66.3 & 78.74$_{-2.01}^{+2.04}$ & 11.97$_{-0.85}^{+0.75}$ & 0.825$\pm$0.009\\
0083250101 & 2001-04-08 & 5158 &  EPIC-pn  & 19.2 & 2.91$_{-0.11}^{+0.10}$ & 0.51$_{-0.25}^{+0.26}$& \ldots\\ 
0144530101 & 2003-05-19 & 5929 &  EPIC-pn  & 67.1 & 8.24$_{-1.03}^{+0.17}$ & 1.61$_{-0.39}^{+0.11}$& \ldots\\ 
0406840301 & 2007-01-17 & 7268 &  EPIC-pn  & 69.2 & 34.76$_{-0.37}^{+0.14}$ & 3.75$_{-0.12}^{+0.17}$& \ldots\\ 
0506220101 & 2008-01-11 & 7627 &  EPIC-pn  & 75.8 & 44.79$_{-0.28}^{+0.52}$ & 5.03$_{-0.18}^{+0.11}$& \ldots\\
0556350101 & 2009-01-31 & 8013 &  EPIC-pn  & 69.6 & 54.71$_{-0.33}^{+0.47}$ & 5.93$_{-0.18}^{+0.11}$& \ldots\\
0601200101 & 2009-12-11 & 8327 &  EPIC-pn  & 76.5 & 62.38$_{-0.27}^{+0.45}$ & 7.09$_{-0.17}^{+0.17}$& \ldots\\
0650420101 & 2010-12-12 & 8693 &  EPIC-pn  & 53.1 & 69.87$_{-0.44}^{+0.54}$ & 8.35$_{-0.27}^{+0.18}$& \ldots\\
0671080101 & 2011-12-12 & 9058 &  EPIC-pn  & 61.4 & 75.40$_{-0.46}^{+0.10}$ & 9.91$_{-0.18}^{+0.18}$& \ldots\\
0690510101 & 2012-12-11 & 9423 &  EPIC-pn  & 60.1 & 78.68$_{-0.62}^{+0.17}$ & 11.21$_{-0.23}^{+0.24}$& \ldots\\
0743790101 & 2014-11-29 & 10141 & EPIC-pn  & 57.8 & 78.00$_{-1.16}^{+0.72}$ & 12.9$_{-0.47}^{+0.20}$& \ldots\\[-3pt]
\enddata
\tablenotetext{\tfn}{\chandra\ ACIS observations with offset chip positions.}
\end{deluxetable}
\twocolumngrid

\section{Analysis and Results}
\label{section:analysis}
\subsection{X-ray Light Curve}
\label{section:lightcurve}
\begin{figure*}[t]
\includegraphics[width=0.345\textwidth]{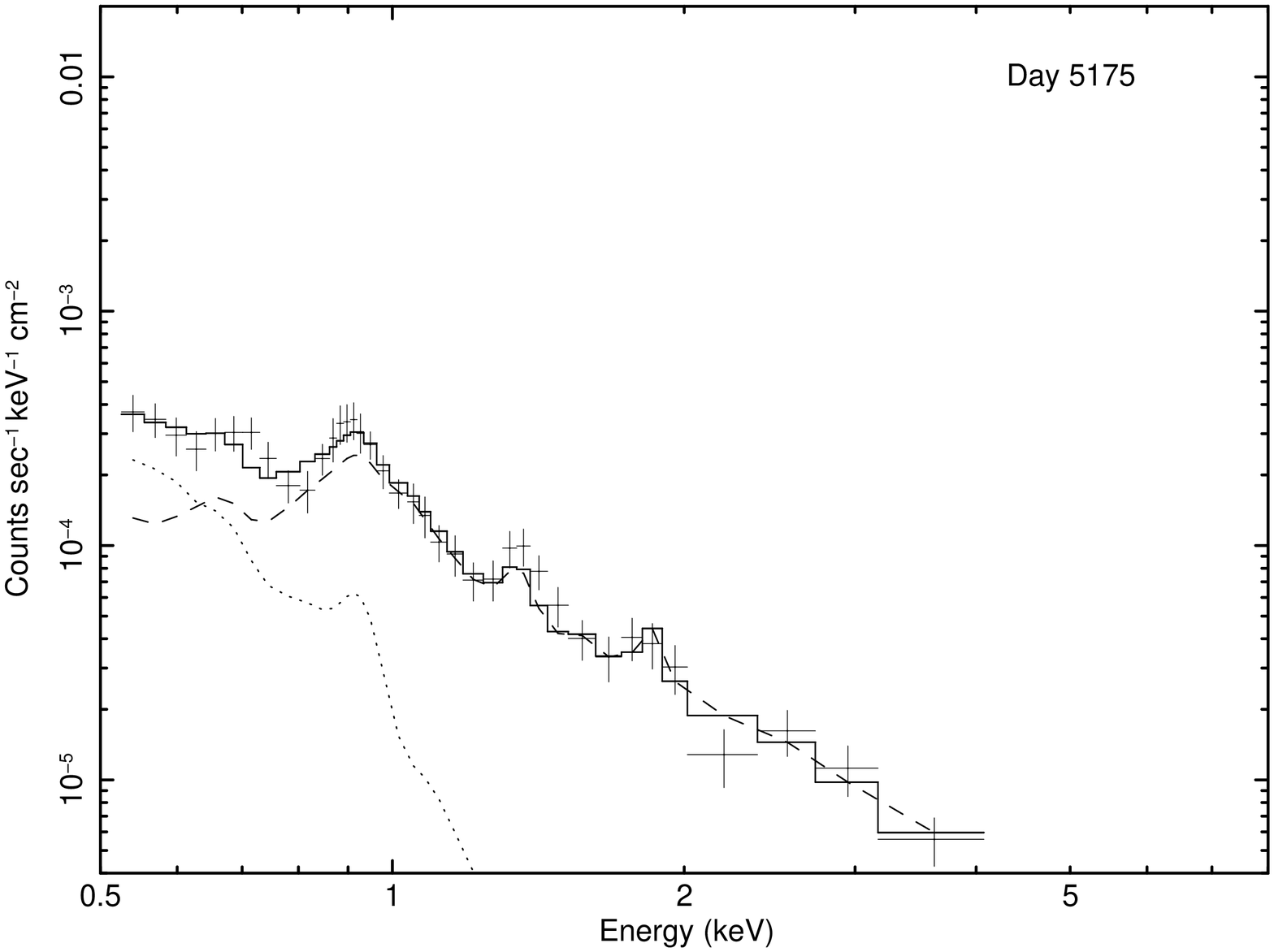}
\includegraphics[trim={0.38in 0.0in 0.00in 0.0in},clip,width=0.33\textwidth]{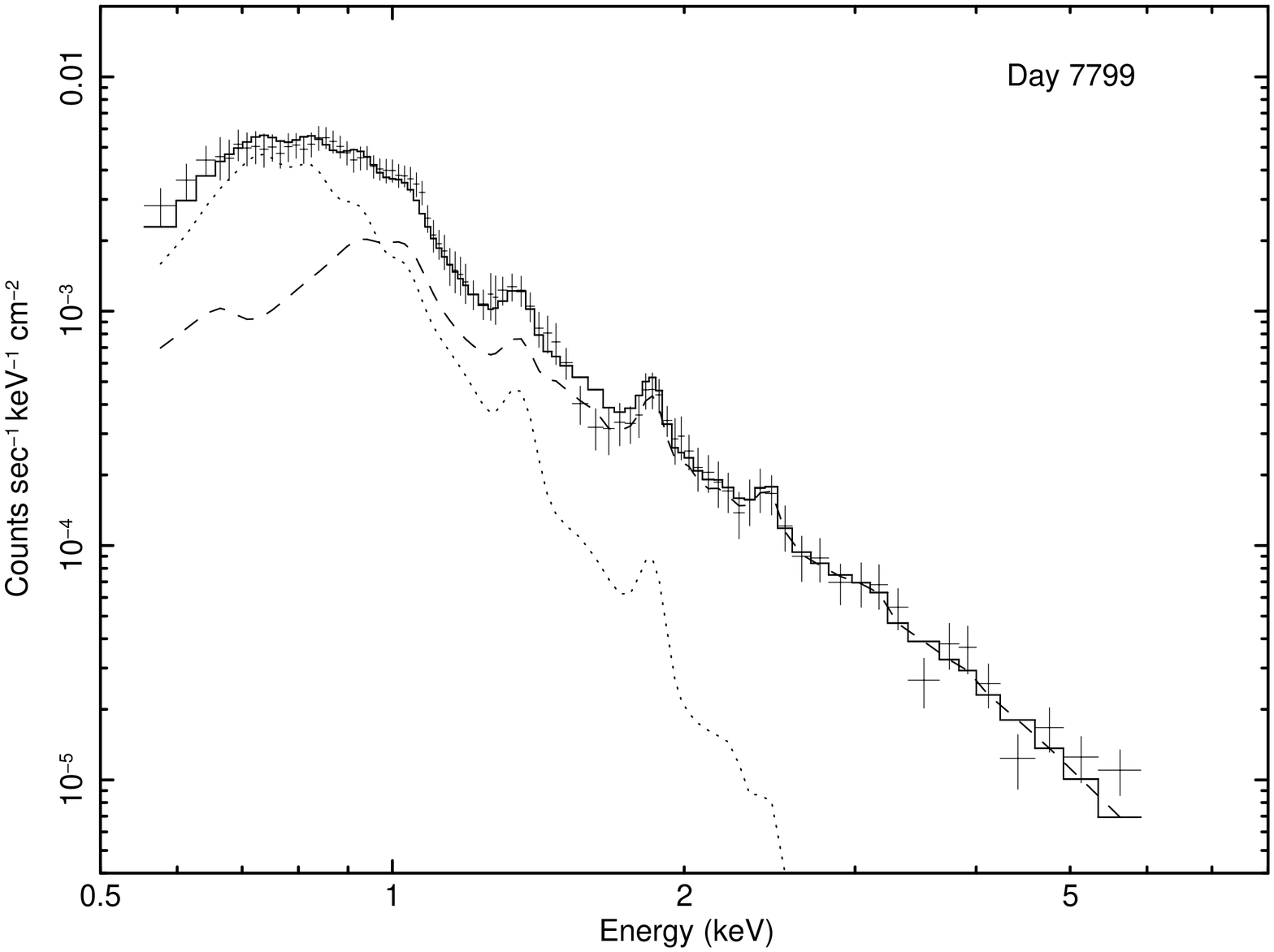}
\includegraphics[trim={0.38in 0.0in 0.00in 0.0in},clip,width=0.33\textwidth]{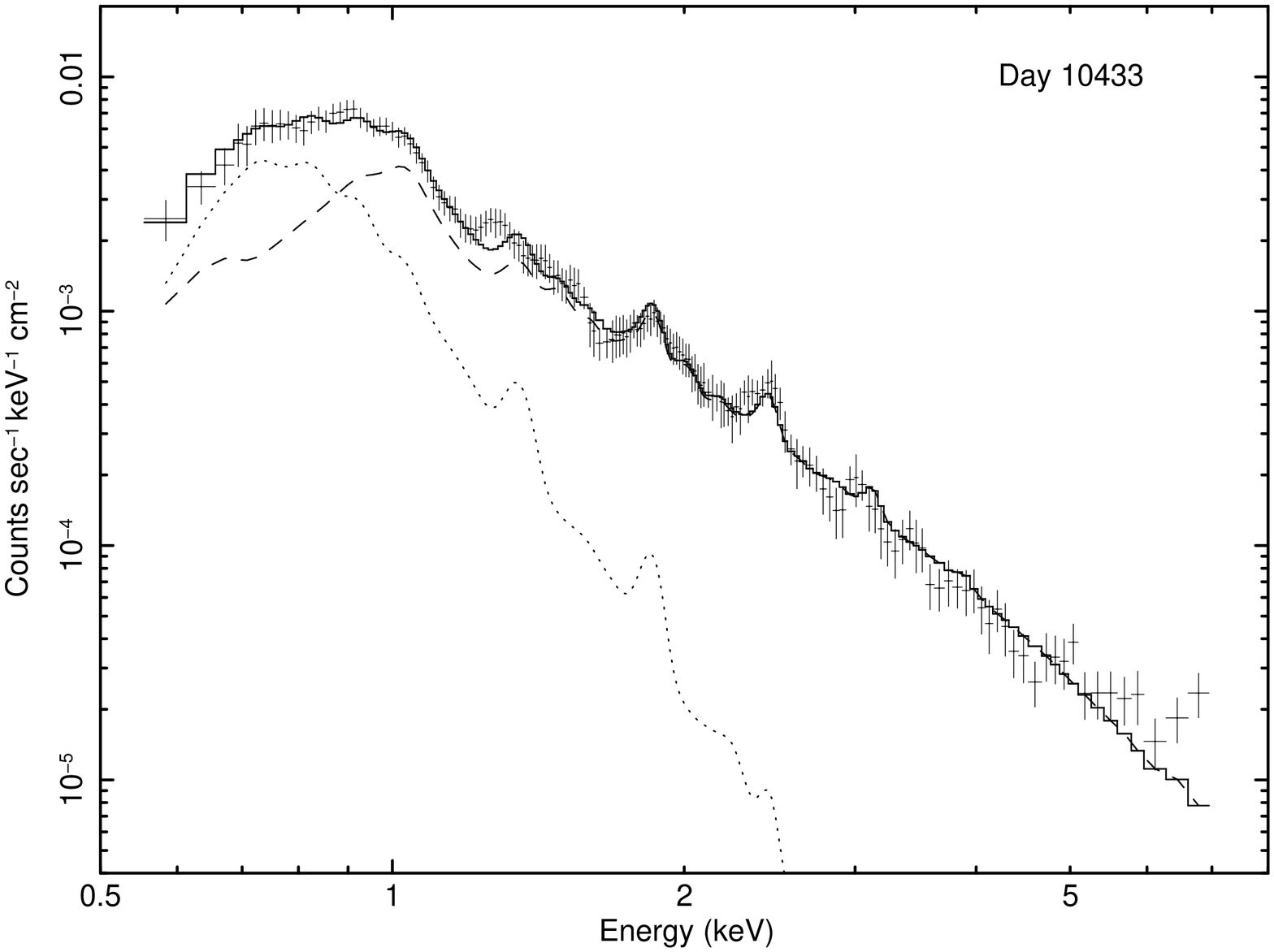}
\caption{\footnotesize ACIS spectra and their best-fit models from days 5175 (left), 7799 (center), and 10433 (right).  The cool ($\sim$0.3 keV) and warm ($\sim$1.8 keV) components are also shown as dotted and dashed lines respectively.}
\label{fig:acisspectra}
\end{figure*}

Each pileup-corrected ACIS spectrum and the LETG spectrum was fit with an absorbed two-component spectral model using \xspec\ 12.8.2 \citep{Arnaud1996}, similar to our previous works.  The model consists of a cool ($\sim$0.3 keV) component in collisional ionizational equlibrium ({\tt vequil}) and a warmer ($\sim$1.8 keV) non-equilibrium component ({\tt vpshock}).  The non-equilibrium model utilized {\xspec} nei version 3.0, which uses the AtomDB 3.0 atomic database. The He and C abundances were fixed to those from the optical analysis of \citet{Lundqvist1996}, Ar, Ca, and Ni to LMC values \citep{Russell1992}, and N, O, Ne, Mg, Si, S, and Fe to those measured by \citet{Zhekov2009} from the deep, high-resolution \chandra\ LETG and HETG observations.  Absorption was fixed to $2.35\times10^{21}$ cm$^{-2}$ \citep{Park2006}.  The temperatures, normalizations, and ionization age were free parameters.  \edit1{The early observation Obs ID 1044, which has the lowest number of counts (1680 compared to 2420-32000 for all others observations), was fit with a single {\tt vpshock} model that is otherwise identical to the described two-component model, as this provided a better fit.}  
For the LETG spectra, all orders were fit simultaneously, and an additional Gaussian smoothing was included in the model. 
Examples of fitted ACIS spectra are shown in Figure \ref{fig:acisspectra}.  The \xmms\ spectra were fit to the same model, but with a second {\tt vpshock} component and free O, Ne, Mg, Si, S, and Fe abundances and absorbing column density.  As \xmms\ has much higher sensitivity than \chandra\ and therefore an order of magnitude more photons, the additional parameters were necessary to obtain fits of sufficient quality for accurate flux measurements.   \edit1{The best-fit absorbing column density is comparable to that used to fit the ACIS spectra, but slightly higher at $\sim3.2\times10^{21}$\nhunit. Abundances are also somewhat higher, by a factor of roughly 1.3-2, except for Fe, which is the same as our \chandra\ measurements. These differences have no significant effect on the measured fluxes, which are the primary concern here; we therefore defer further investigation to a future work.} 
The \xmms\ spectral model is comparable to that used in \citet{Maggi2012}, and we obtained similar fluxes.

An important consideration in measuring the fluxes is the growing molecular contamination on the ACIS optical blocking filter (OBF) \citep{ODell2013}.  This contamination partially absorbs the flux at energies $\lesssim$2 keV, and is strongest at energies below 1 keV. 
It has evolved unpredictably, and as a result it has sometimes been difficult to obtain accurate contamination models until many months after a given observation. 
In the past, this has led to inaccurately low \sna\ flux measurements \citep{Park2011,Helder2013}.  In addition, the buildup of contamination increases with distance from the chip center, and several of the \sna\ monitoring observations used offcenter positions (to minimize pile-up by reducing the frame readout time).  The contamination models are more poorly calibrated for these positions, and the measured fluxes for these observations (noted in Table \ref{table:observations}) should be interpreted with an extra degree of caution. 

We have taken several steps to mitigate contamination issues as much as possible.  The most up-to-date contamination models available, version 9 released in 2014 July, have been used for the ACIS spectra.  Additionally, we have obtained several independent, uncontaminated fluxes for comparison.  The \chandra\ HRC detector does not suffer from contamination, so Obs ID 16757 from 2015 March utilized the HRC-S/LETG instrument configuration to obtain high-resolution spectra, from which we acquired the uncontaminated 0.5-2 keV flux.  We also obtained fluxes from the \xmms\ EPIC-pn observations of \sna\, through 2014 November.  
Given the agreement between the ACIS, HRC-S/LETG, and \xmms\ fluxes, we believe the contamination has been adequately taken into account, but it should be kept in mind that if the  contamination is significantly worse than expected for the most recent observation, the associated ACIS soft flux may be slightly underestimated.  

\begin{figure*}[htbp]
\includegraphics[trim={0.3in 0.2in 0.5in 0.1in},clip,width=0.5\textwidth]{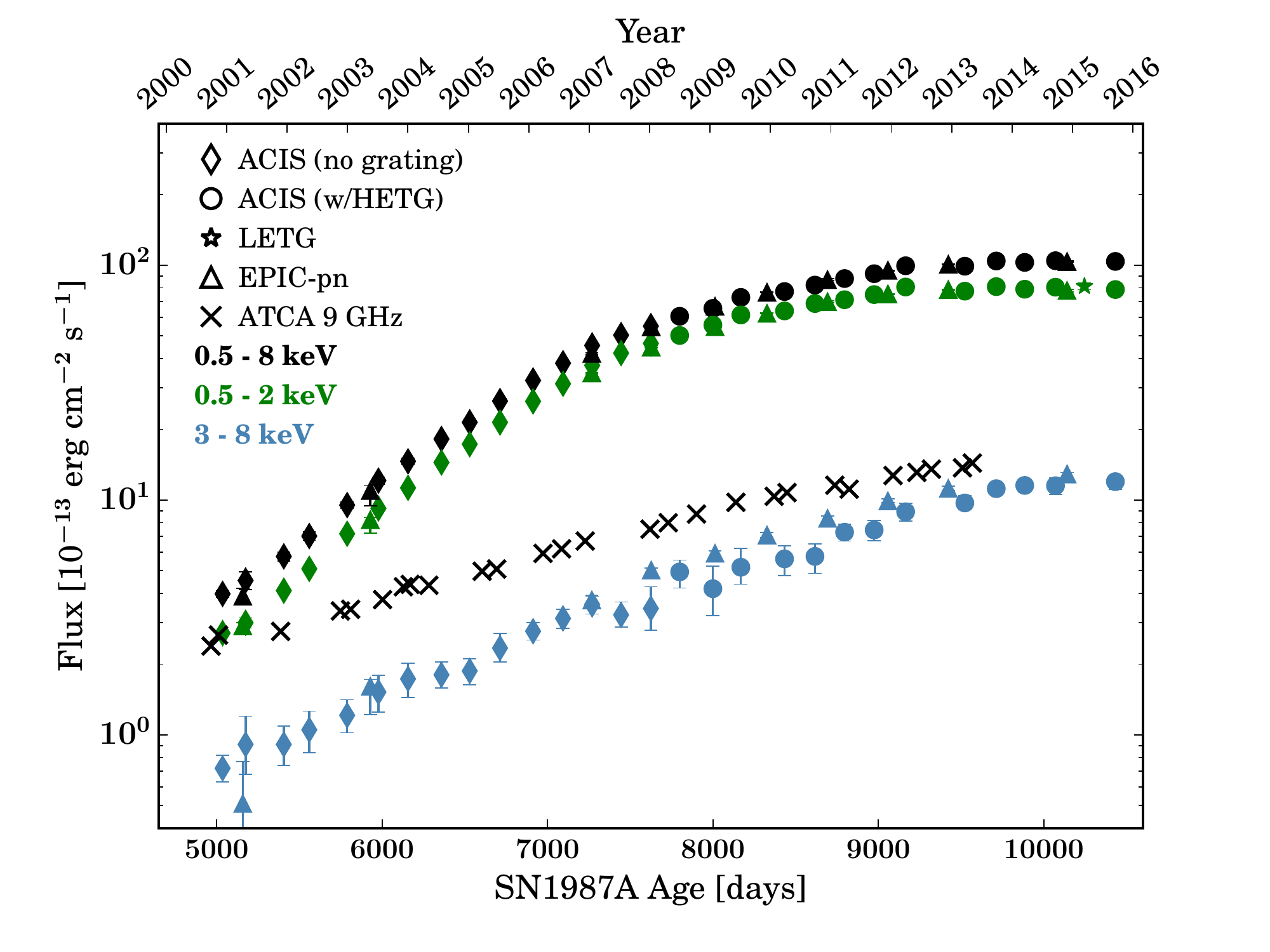}
\includegraphics[trim={0.3in 0.2in 0.5in 0.1in},clip,width=0.5\textwidth]{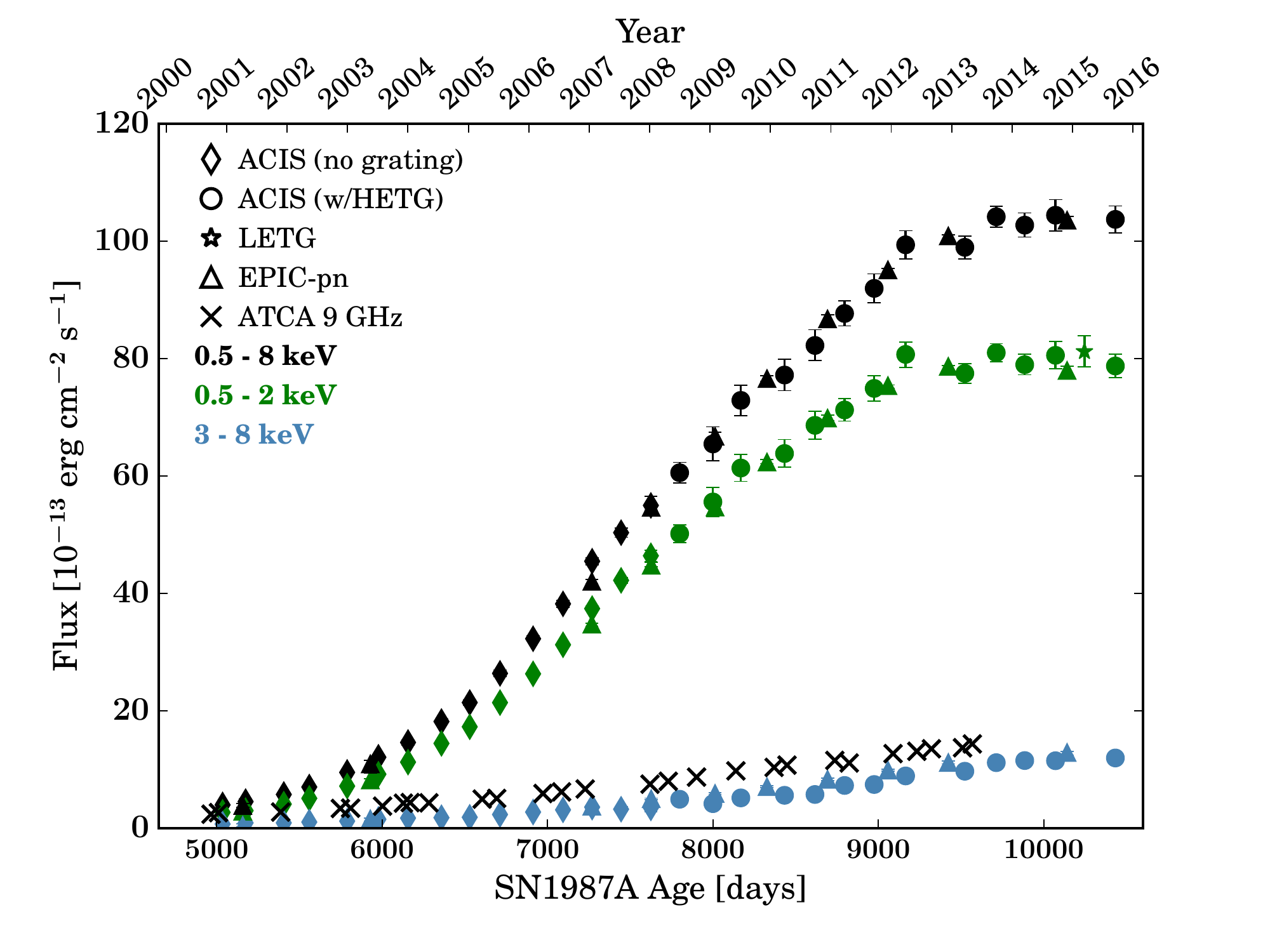}
\caption{\footnotesize X-ray light curve of \sna\ from days 5036 through 10433, shown with fluxes on both a log (left) and linear (right) scale.  \chandra\ \acis\ fluxes are given as diamonds (bare \acis\ observations) and circles  (\acis\ observations with HETG), and the stars are LETG fluxes from the 2015 March observation.  \xmm\ EPIC-pn fluxes are shown as triangles.  The 0.5-8.0 keV fluxes are in black, 0.5-2.0 keV in green, and 3.0-8.0 keV in blue.  The 9 GHz ATCA fluxes from \citet{Ng2013}, arbitrarily scaled, are shown as crosses.  Note that for many points the error bars are too small to be visible.}
\label{fig:lightcurve}
\end{figure*}

The resulting 0.5-2.0 keV and 3.0-8.0 keV fluxes are provided in Table \ref{table:observations} and the light curves in Figure \ref{fig:lightcurve}. 
 As described in previous works, there is a sharp upturn in the soft light curve around 6000 days, due to the blast wave impacting the main body of the ER \citep{Park2005}. Between 7000 and 8000 days the light curve changes again, such that the flux increases linearly rather than exponentially \citep{Helder2013}.  The latter change indicates the average density of new material encountered by the blast wave stopped increasing around that time.  The soft \edit1{band flux (0.5-2 keV)} since day 9500 has remained approximately constant; such a leveling off of the light curve is expected to happen when the blast wave leaves the dense ring \citep[e.g.][]{Park2011}.  In contrast, the hard light curve has overall exhibited slower growth with fewer changes over time, increasing steadily until at least day $\sim$10000.

\begin{figure}[htbp]
\includegraphics[trim={0.2in 0.2in 0.5in 0.1in},clip,width=1.05\columnwidth]{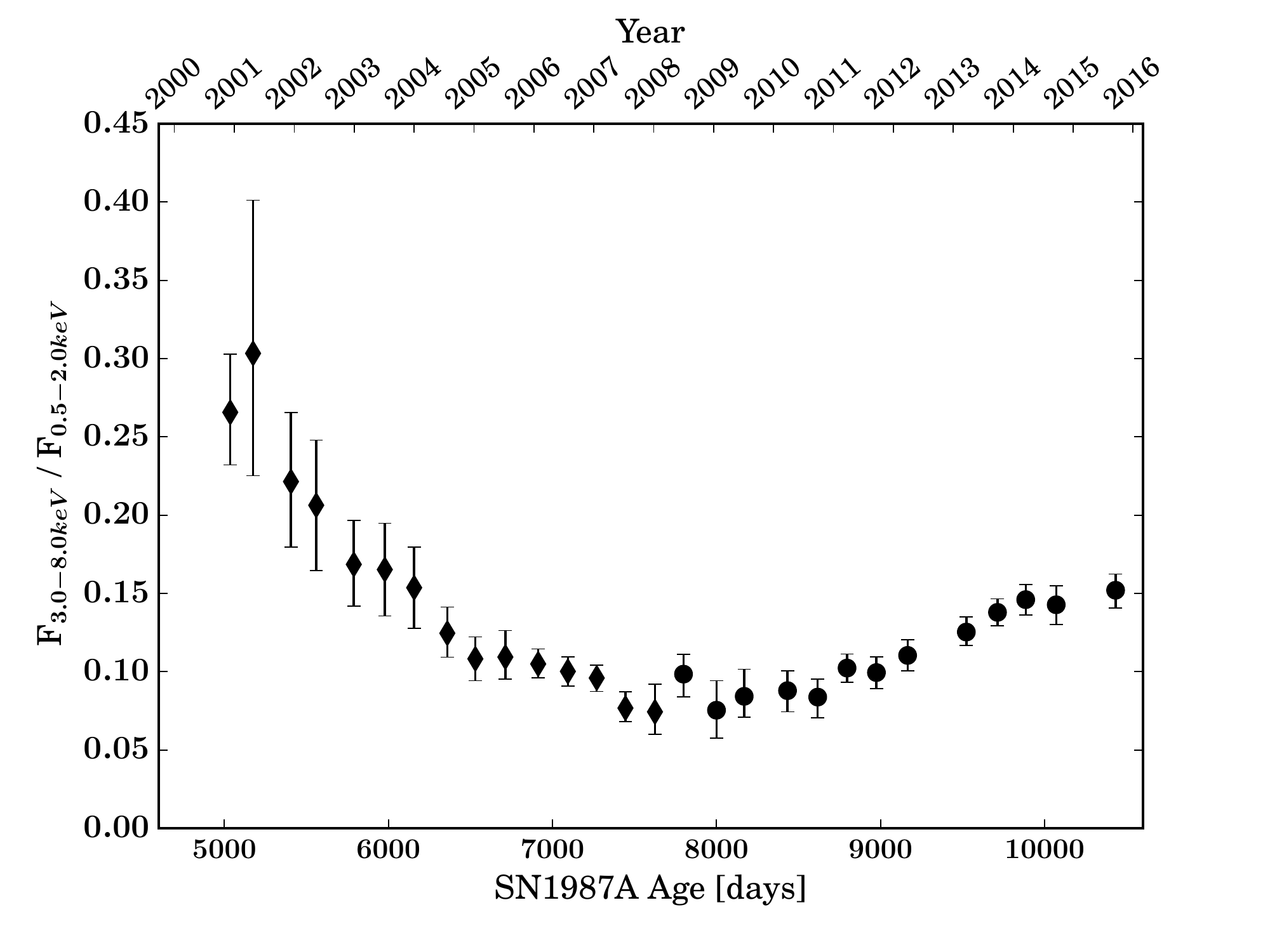}
\caption{\footnotesize Ratio of the 3-8 keV to 0.5-2 keV ACIS fluxes from days 5036 to 10433.  The symbols are the same as in Figure \ref{fig:lightcurve}.}
\label{fig:hardnessratio}
\end{figure}

\begin{figure}[htbp]
\includegraphics[trim={0.2in 0.2in 0.5in 0.1in},clip,width=1.05\columnwidth]{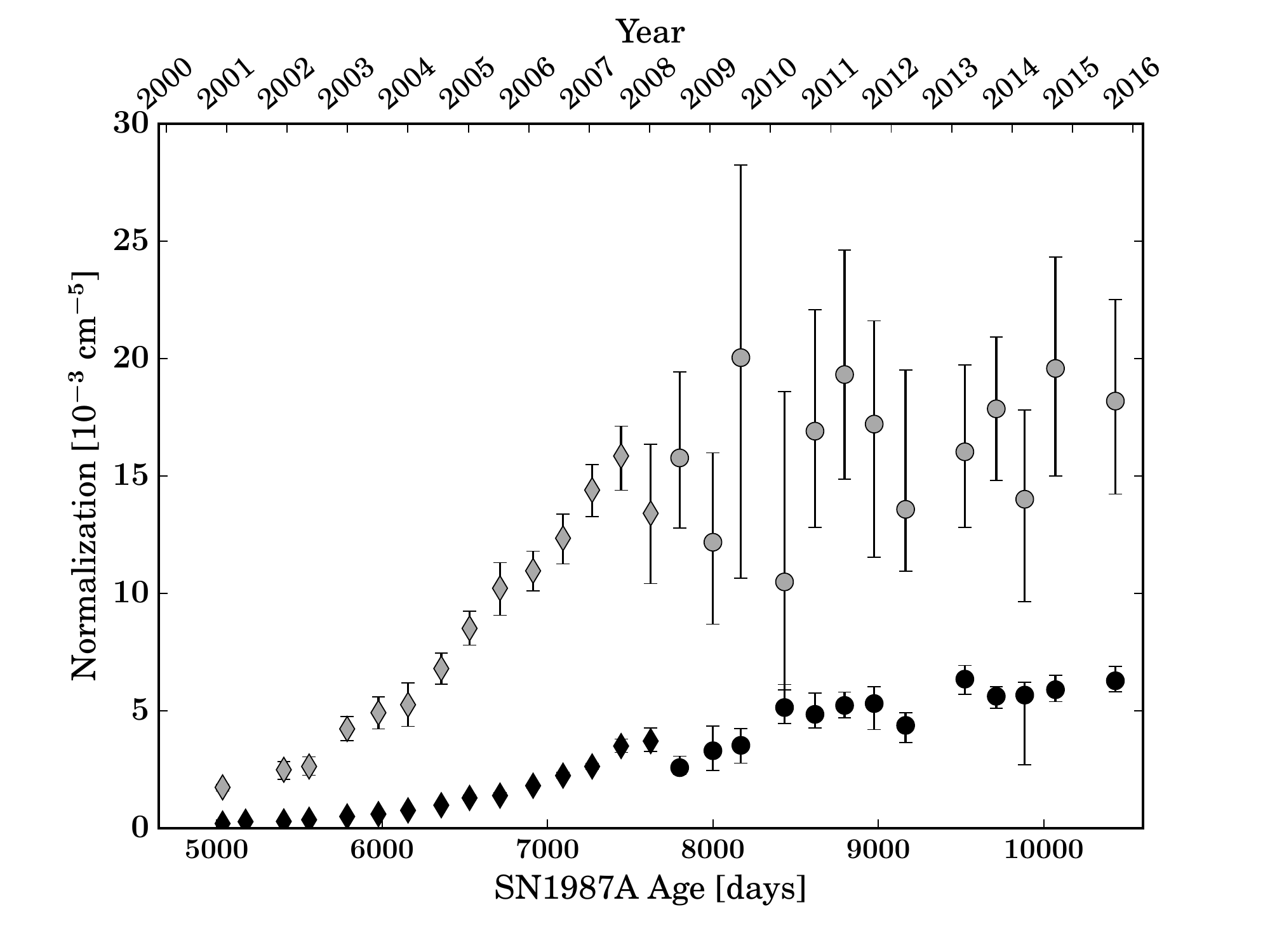}
\caption{\footnotesize Normalizations of the cool (gray) and warm (black) spectral model components from days 5036 to 10433.  The symbols are the same as in Figure \ref{fig:lightcurve}.}
\label{fig:norms}
\end{figure}

The changes in the light curves are also reflected in the ratio of the 3.0-8.0 keV to 0.5-2.0 keV band fluxes \edit1{and the normalizations of the cool and warm components of the spectral model, shown in Figures \ref{fig:hardnessratio} and \ref{fig:norms}. The model normalization parameters represent the scaled volume emission measures.} The X-ray emission continuously softens until about day 7500, when both the band ratio \edit1{and the cool component normalization abruptly flatten.  The band ratio then} begins to slowly increase, with a possible flattening in the last 2 or 3 observations.


\subsection{Imaging}
\label{section:imaging}
\begin{figure*}[htbp]
\includegraphics[width=\textwidth]{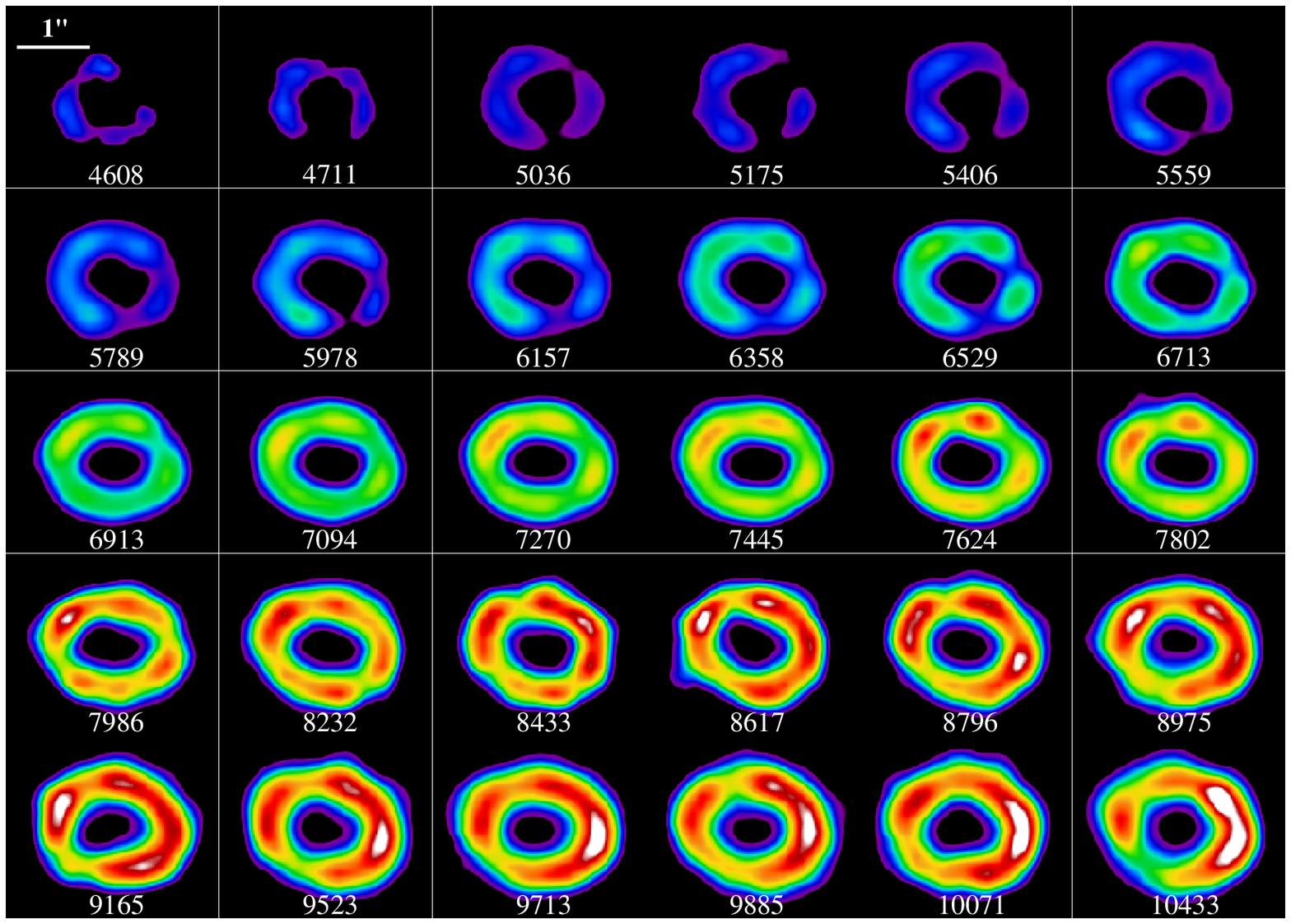}
\caption{\footnotesize Deconvolved, smoothed 0.3-8.0 keV \edit1{false-color} images of \sna\ covering days 4608 - 10433.  Images use a square root scale and are normalized by flux.  The age, in days since the supernova, is shown below each image. North is up and East is to the left.}
\label{fig:images}
\end{figure*}
\begin{figure}[htbp]
\includegraphics[trim={0.2in 0.2in 0.5in 0.1in},clip,width=\columnwidth]{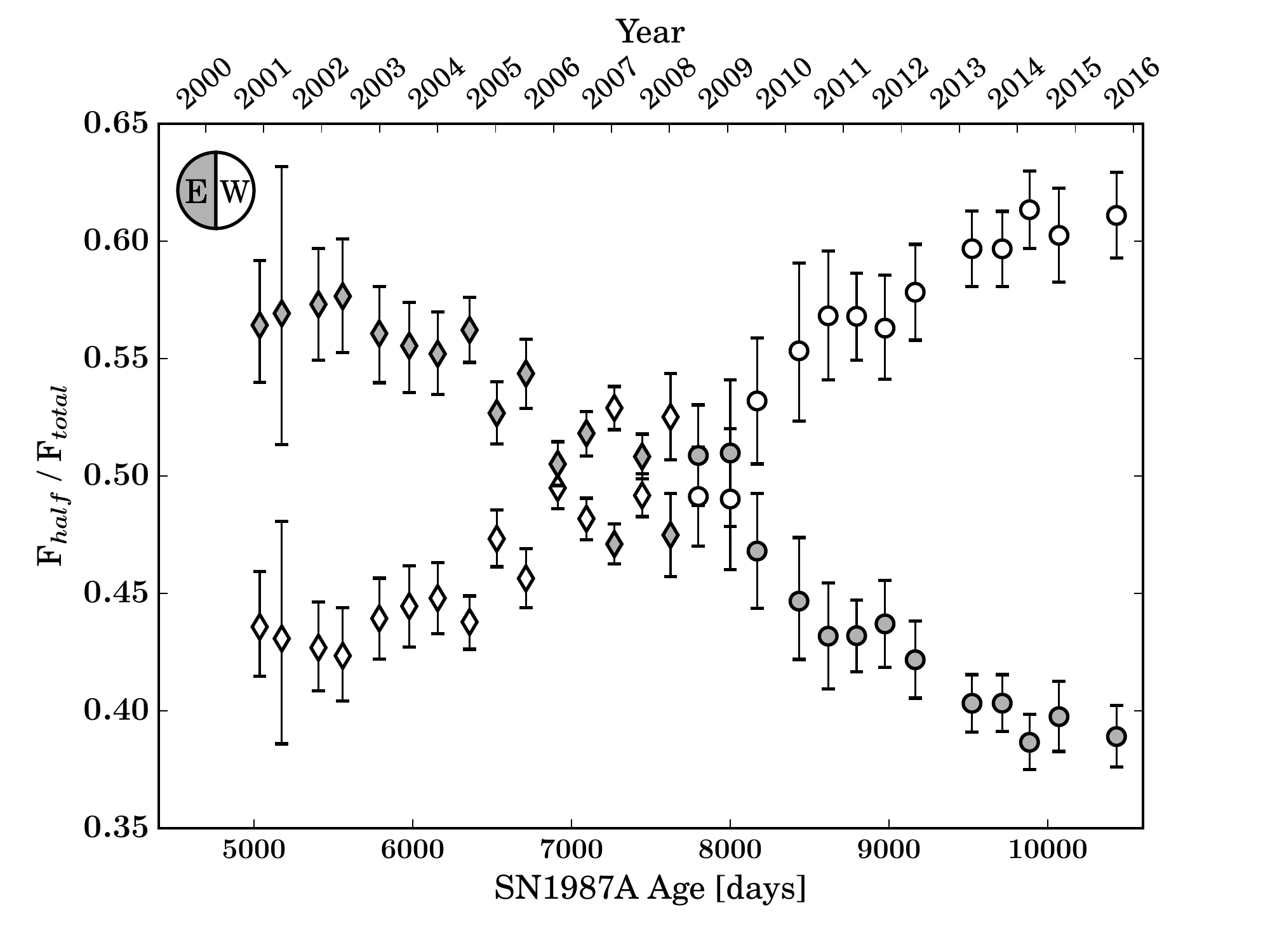}
\caption{\footnotesize Fraction of the total 0.3-8 keV flux in the east (filled symbols) and west (empty symbols) halves of the ER over time.  The center of the ring for each observation is defined as the center of the ring from our best-fit model as described in \S\ref{section:expansion}.  Symbols are the same as Figure \ref{fig:lightcurve}.  The fractional fluxes of the individual southeast and northeast quadrants evolve similarly over time (both decreasing), as do the western quadrants (both increasing).}
\label{fig:fluxfractions}
\end{figure}

Images from all our \chandra\ epochs are shown in Figure \ref{fig:images}. Initially brightest in the east, the ER becomes obviously brighter in the west by 8433 days and has remained that way through the most recent observation.  A plot of the fraction of the total flux in the east and west halves over time demonstrates that reversal of the asymmetry occurred between 7000 and 8000 days (Figure \ref{fig:fluxfractions}). The most recent image at 10433 days suggests the southeastern quadrant of the ER is beginning to fade out, while the west remains bright.

\subsection{Expansion}
\label{section:expansion}
\begin{figure*}[htbp]
\includegraphics[trim={0.2in 0.2in 0.5in 0.1in},clip,width=0.5\textwidth]{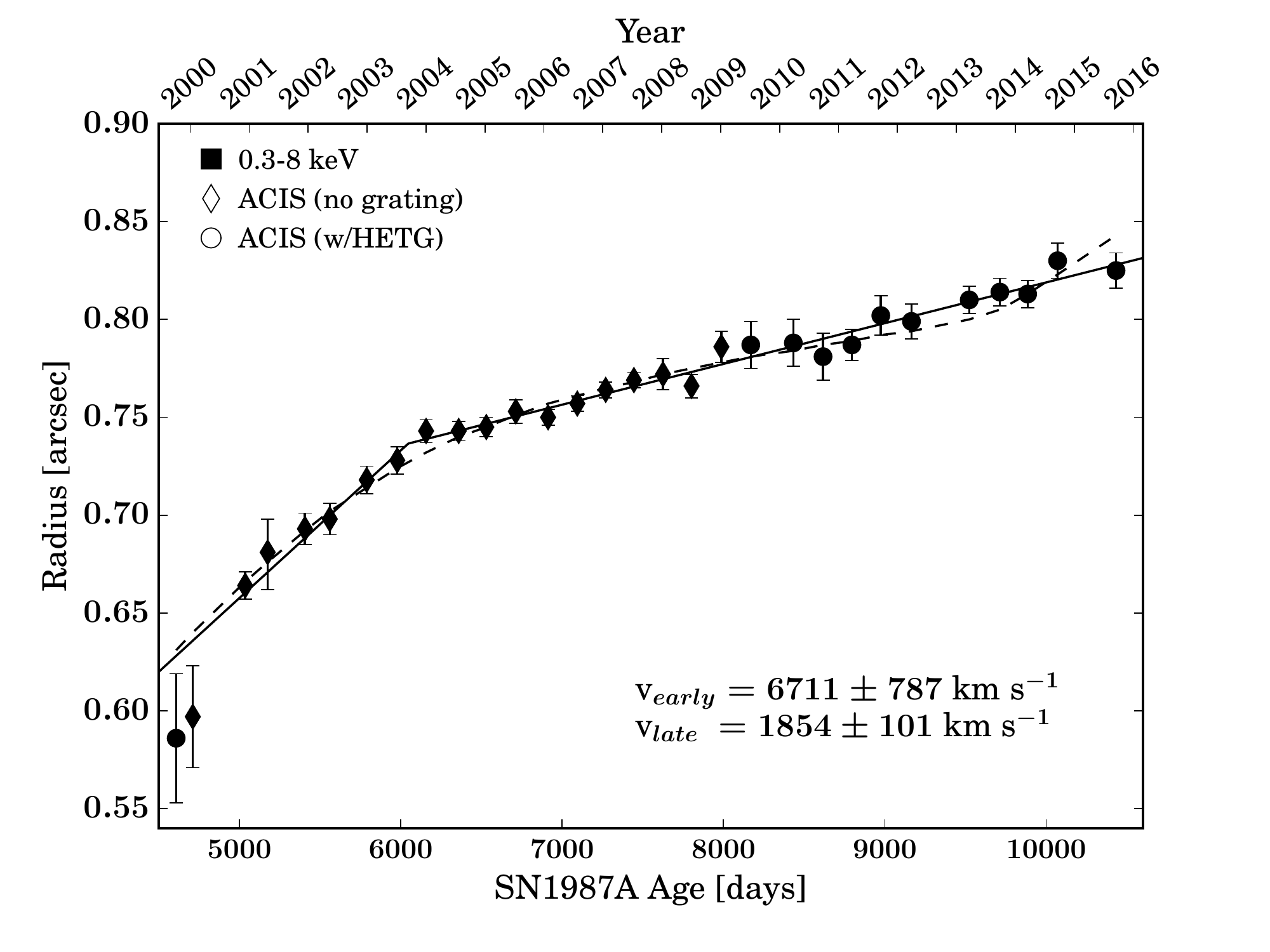}
\includegraphics[trim={0.2in 0.2in 0.5in 0.1in},clip,width=0.5\textwidth]{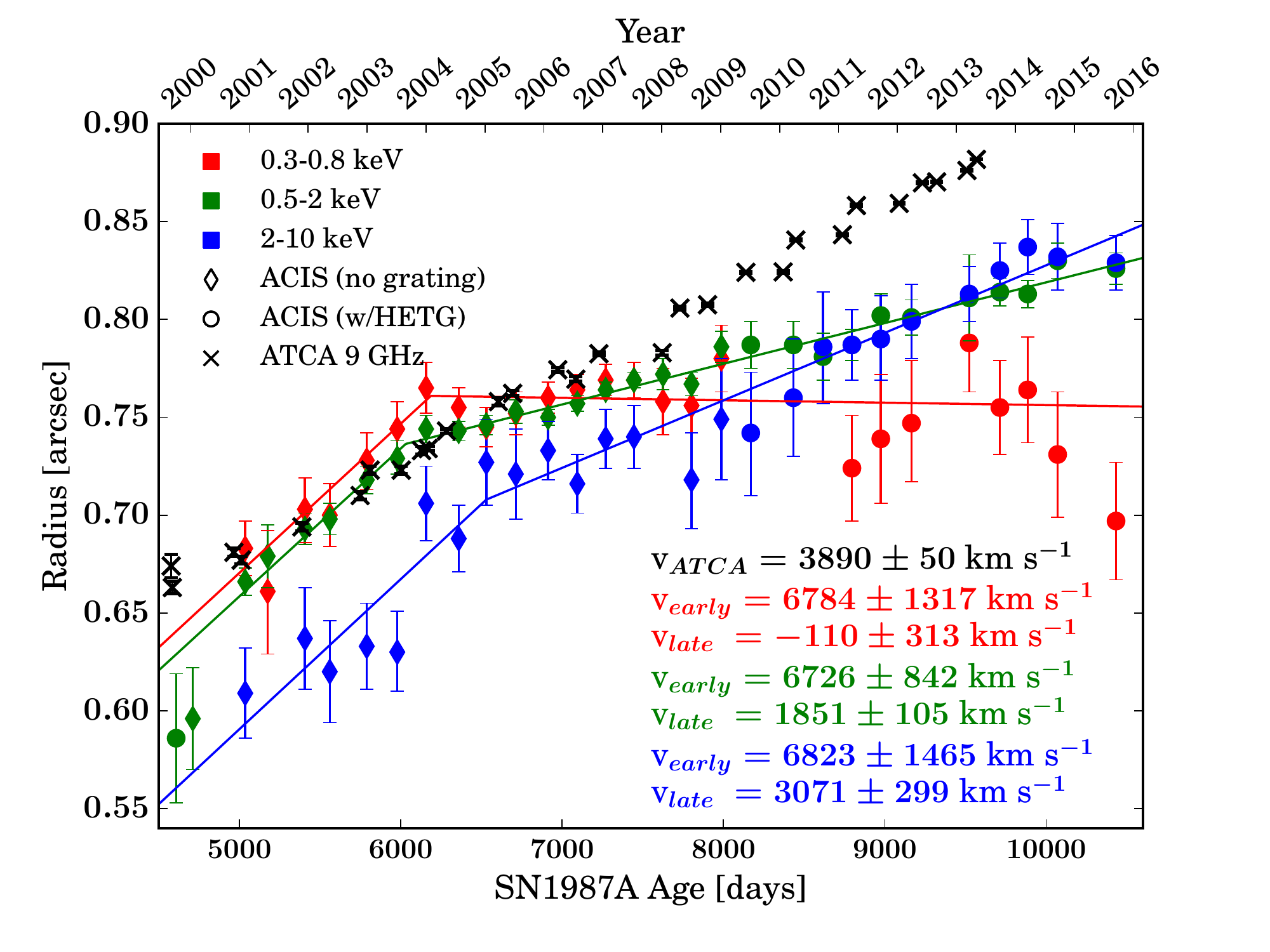}
\caption{\footnotesize Best-fit radii of the ER from the ring plus 4-lobe model fits. On the left is the result from fitting to the broad band (0.3-8 keV) images. On the right are the results to fitting to the 0.3-0.8  keV (red), 0.5-2 keV (green), and 2-10 keV (blue) images.  The crosses show the semi-major axis of the ER at 9 GHz from \citet{Ng2013}. A simple broken-linear fit to each band is also shown as a solid line of the corresponding color, with the resulting velocities \edit1{with 1$\sigma$ errors} noted for the early and late epochs (before and after $\sim$6000 days, respectively).  \citet{Ng2013} found the radio expansion was better fit with a simple linear model; the velocity from their linear fit is also given. \edit1{In the left panel, an analytical density profile was used to fit the 0.3-8 keV expansion, as discussed in \S\ref{section:discussion}, and the resulting fit is shown as a dashed line.}}
\label{fig:expansion}
\end{figure*}

The superb spatial resolution of \chandra\ enables us to measure the radial expansion of the ER.  This was done following the method of \citet{Racusin2009}, wherein each deconvolved image was fit to a spatial model that consists of four lobes and a ring.  The best fit radii are shown in Figure \ref{fig:expansion}.  We carried out this procedure for images in the 0.3-8 keV, 0.3-0.8 keV, 0.5-2 keV, and 2-10 keV bands.  The resulting best fit radii for the 0.3-8 keV band are given in Table \ref{table:observations}. A simple broken-linear function was fit to the results for each band to estimate expansion velocities.  The statistics are worse for the 0.3-0.8 keV and 2-10 keV bands due to the lower number of counts in these bands for many of the observations, especially after insertion of the HETG. This is reflected in the substantially larger error bars. For each band, observations which had insufficient counts ($\lesssim$few hundred) for robust image fitting were excluded from the expansion analysis.

The time of impact with the main ER and the velocities before and after this event were determined from the fit to the 0.3-8 keV images. Impact with the ER occured at 6047$\pm$110 days.  The velocities (i.e. the slopes of the expansion curve) are 6711$\pm$787 km s$^{-1}$ before this date and 1854$\pm$101 km s$^{-1}$ afterward.  The early velocity decreases slightly to 6104$\pm$849 km s$^{-1}$ if the earliest two observations, with large error bars, are excluded.  These results are consistent with the earlier estimates of \citet{Racusin2009} and \citet{Helder2013} using the same method.

It is also informative to compare the expansion of the ER in multiple bands.  The standard `soft' band of 0.5-2 keV contains most of the counts, and therefore appears essentially the same as the 0.3-8 keV expansion.  The softest emission, 0.3-0.8 keV, which corresponds to emission from the densest regions, is similar to the broad band, both in the ring size and expansion velocity, until impact with the ER at $\sim$6000 days.  After this date, expansion in this band \edit1{becomes consistent} with zero.  The hard band (2-10 keV) appears to have a slightly later break point, at 6530$\pm$352 days.  Prior to this it has a similar velocity to the other bands but smaller radii.  After ER impact, the velocity is faster than the softer bands, 3071$\pm$299 km s$^{-1}$, and similar to the radio expansion velocities, $\sim$3900 km s$^{-1}$ \citep{Ng2013,Zanardo2013}. Between 8000 and 9000 days the radii in the hard band overtake those of the 0.3-0.8 keV band and catch up to the 0.5-2 keV band.  Currently, the size of the ER in the hard band is similar to that in the 0.3-8 keV and 0.5-2 keV bands, while the 0.3-0.8 keV radius is significantly smaller.  \edit1{These energy-dependent expansion rates are a likely a result of the shock velocity being slower through denser material.}

\section{Discussion}
\label{section:discussion}
The ER can be modeled as a smooth ring with n$\sim$10$^3$cm$^{-3}$ and very dense clumps, n$\sim$10$^4$cm$^{-3}$, distributed around the inner edge of the smooth ring \citep{Dewey2012,Orlando2015}. The X-ray emission arises from a complex system of transmitted and reflected shocks as the blast wave interacts with these various CSM components \citep{Zhekov2009,Zhekov2010,Dewey2012}.  While the actual physical picture is quite complicated, for the purposes of interpreting the X-ray observations the emission can be characterized by a `cool' ($\sim$0.3 keV), or soft, component and a `warm' ($\sim$1.5-3 keV), or hard, component.  The cool component represents the slow transmitted shocks in the dense clump material and is responsible for the majority of the soft X-ray emission after $\sim$6000 days \citep{Zhekov2010,Dewey2012,Orlando2015}.  The warm component is representative of the shocks moving through the lower density ring material, including reflected shocks, and contributes most of the hard emission \citep{Zhekov2010,Orlando2015}.

Between days 6000 and 7000 the steep increase of the soft X-ray light curve is due to the blast wave interacting with the dense clumps.  The resulting transmitted shocks in the clumps moved more slowly than the shocks moving through the inner \HII\ region or the smooth component of the ER. This can be seen in the X-ray expansion, shown in Figure \ref{fig:expansion}, \edit1{of the 0.3-0.8 keV band. This emission is dominated} by the densest clump emission which slowed the transmitted shock dramatically, such that the measured expansion velocity is consistent with zero after day 6000.  Expansion in the 0.5-2 keV band, which contains most of the X-ray counts \edit1{and includes contributions from both the clumps and smooth component of the ring,} also slowed significantly after day 6000, to a velocity of 1854 km s$^{-1}$.  This velocity is consistent with that derived by \citet{Dewey2012} from hydrodynamical modeling of the high-resolution X-ray spectra.  
\edit1{We can also estimate the density jump required to cause the observed deceleration as the blast wave moved from the \HII\ region into the ER.  \citet{Zhekov2010} assumed an analytical density profile that provides an analytical solution to the expansion curve \citep[see \S3.1 of][]{Zhekov2010}. We fit this model to the 0.3-8 keV expansion curve; the result is shown in Figure \ref{fig:expansion} (left). Our best-fit model requires that the density increases by a factor of $10.6\pm1.6$. Given that the \HII\ region has densities of $\sim10^2$cm$^{-3}$, this indicates that the the expansion is primarily related to the smooth ring component, with a typical density on the order of $10^3$ cm$^-3$.
For} this epoch, from days 6500 - 8000, \citet{Dwek2010} found the infrared-to-X-ray band \edit1{flux} ratio was approximately constant, implying X-ray heating of dust in the ER.  Throughout this period, the optical flux continuously increased \citep{Fransson2015} as transmitted shocks moved through the clumps.

The hard X-ray light curve is dominated by emission from the smooth component and so increases more slowly, resulting in the sharp decline of the band ratio (Figure \ref{fig:hardnessratio}) before day 7500.  This behavior is also evident in our 2-component spectral fits, shown in Figure \ref{fig:acisspectra}.  Until $\sim$7500 days the cool component increased much more rapidly than the warm component (Figure \ref{fig:norms}).  The radio light curve and expansion resemble those of the hard X-rays (Figures \ref{fig:lightcurve} and \ref{fig:expansion}).  
The similar evolution of the hard X-rays and the radio suggests the radio emission also originates from the smooth ring component.

After $\sim$7500 days, both the optical and mid-infrared emission from the ER began to fade \citep{Fransson2015,Arendt2016}, which is interpreted as the result of destruction of the clumps and dust, respectively.  Around this time, we find the growth of the soft X-ray light curve transitions from exponential to linear (Figure \ref{fig:lightcurve}) and the band ratio flattens (Figure \ref{fig:hardnessratio}), which suggest there is little emission from newly shocked clump material.  
We can use this to place some rough limits on the lengths of the clumps, which likely have a range of different sizes, radial distances, and shock velocities. The largest possible clump (i.e. the longest protrusion) would have been encountered by the forward shock between 4000 and 5000 days (coincident with the appearance of the first hot spots), with the resulting transmitted shock traversing the entire clump by day 8000 (the approximate end of shock-clump interaction).  For a velocity of 1854 km s$^{-1}$, as found in our expansion measurements, this implies a length of $\sim$6.4$\times10^{16}$cm.  The smallest clumps would have been encountered by day 6045 (onset of full interaction with the ER) with the shocks exiting around day 7000 (the first signs of decreasing shock-clump interaction), implying a length of $\sim$1.5$\times10^{16}$cm.  This is comparable to the 1.7$\times10^{16}$cm clump size estimated by \citet{Orlando2015} with their simple clump model.

During the post-clump phase, shocks were still moving through the smooth ring component and thus both the hard X-ray light curve and expansion continue to increase as before. Comparing the ACIS spectra at days 7799 and 10433 (Figure \ref{fig:acisspectra}), it is clear that the cool component did not increase, while the warm component has continued to grow. This explains the observed flattening of the band ratio at $\sim$7500 days (Figure \ref{fig:hardnessratio}) and the slow increase afterwards as soft emission from the clumps very slowly begins to fade while the soft and hard emission from the smooth component both continue to increase at a steady rate.  \edit1{There is no detectable change in the expansion velocity, as expected if the observed expansion is mainly that of the shock moving through the smooth ring.}

The soft X-ray light curve leveled off at $\sim$$8\times10^{-12}$ ergs s$^{-1}$cm${^{-2}}$ by day 9500.  Such a flattening of the light curve has been predicted to occur when the forward shock finally leaves the ER \citep{Park2011,Dewey2012,Orlando2015}.  In this same period new optical emission was seen by \citet{Fransson2015} outside of the ER, as both faint hot spots and diffuse emission.  This is interpreted as gas that is either directly shocked by the blast wave or heated by X-rays from the outer edge of the ER.  In either case, this lends support to the idea that the forward shock is now beginning to propagate into the region beyond the ER.  

While the X-ray emission has thus far been dominated by the shocked \HII\ material (prior to day 4000) and then the shocked ER material, the simulations of \citet{Orlando2015} determined that the flux from the reverse-shocked ejecta was also steadily growing.  As the reverse shock continues to encounter more ejecta and the emission from the ER begins to fade, emission from the ejecta will become the dominant source of X-rays.  \citet{Orlando2015} predict this transition will occur within the next five years.

The evolution of X-ray morphology, dominated by soft emission, is similar to the optical. The optical hot spots and the X-ray emission first appeared on the eastern side of the ER \citep{Sugerman2002,Fransson2015}, and the radio emission is also stronger in the east \citep{Ng2013,Zanardo2014}.  \citet{Fransson2015} found that the hot spots on the eastern side began to fade around day 7000, while those in the west did not begin to fade until $\sim$1500 days later.  The asymmetry in the X-ray emission reversed during this time, becoming  brighter in the west by about 8000 days, as seen in Figures \ref{fig:images} and \ref{fig:fluxfractions}.  By day 10433, X-ray emission in the east has clearly started to weaken, especially in the southeast quadrant where the optical emission has also faded the most. 

\begin{figure*}[t]
\includegraphics[trim={0.0in 0.2in 0.0in 0.0in},clip,width=\textwidth]{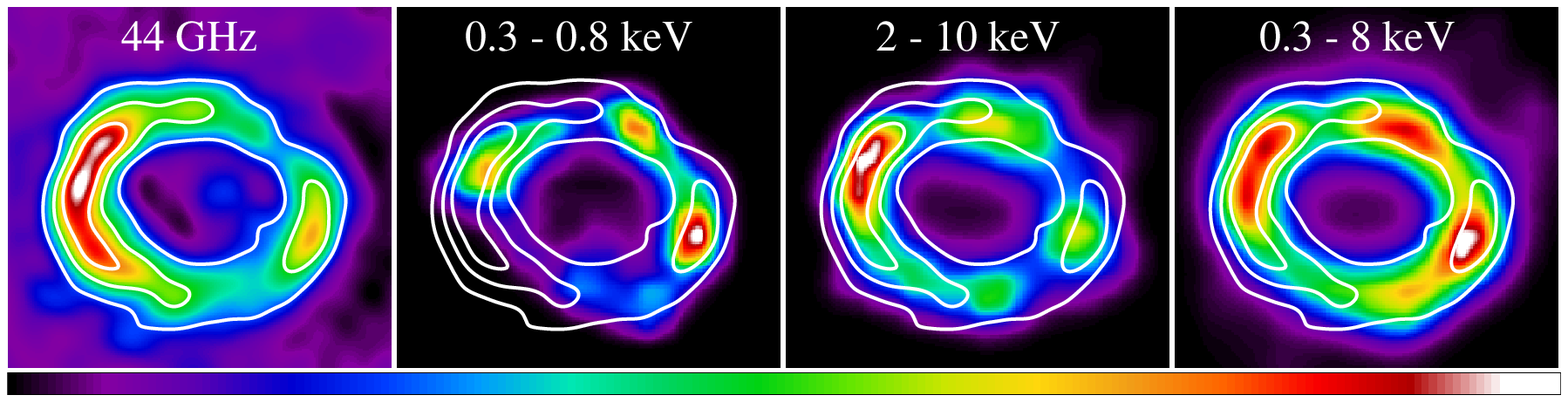}
\caption{\footnotesize From left to right, the 44 GHz \citep[from][]{Zanardo2013}, 0.3-0.8 keV, 2-10 keV, and 0.3-8 keV images, with the 44 GHz contours overlaid in white.  All images are from 2011 March ($\sim$8800 days).  There is good agreement between the broad band X-ray and the radio emission, while the hard X-rays match the radio better than the very soft 0.3-0.8 keV band.}
\label{fig:xrayradioimage}
\end{figure*}

\begin{figure}[h]
\includegraphics[trim={0.0in 0.2in 0.0in 0.0in},clip,width=\columnwidth]{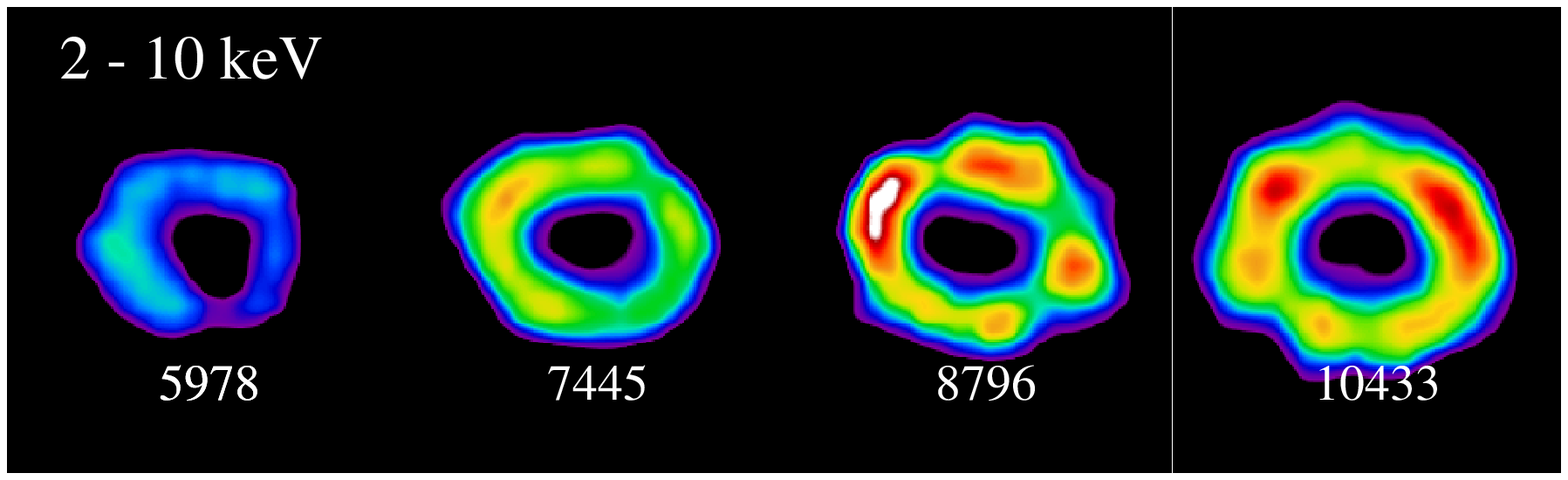}
\caption{\footnotesize 2-10 keV images at roughly 4 year intervals, with the age in days shown underneath.  Prior to $\sim$9500 days, the 2-10 keV images always peak in the east (to the left), with a slow decrease in this asymmetry over time. Since 9500 days, it has been more symmetric and may now be slightly brighter in the west.  Compare this to the broad band images (dominated by softer emission), which are clearly brightest in the west by day 8433. (Figures \ref{fig:images} and \ref{fig:fluxfractions}).}
\label{fig:hardimages}
\end{figure}

As with the light curve and expansion, the radio morphology matches that of the hard X-rays better than the soft (Figure \ref{fig:xrayradioimage}).  They evolve similarly to the optical and the overall X-ray, but are delayed by roughly 2000 days.  In the 2-10 keV band, the east-west asymmetry began to reverse only around day 9500 (Figure \ref{fig:hardimages}), compared to $\sim$7500 days for the 0.3-8 keV emission. \citet{Ng2013} found similar behavior for the 9 GHz images, which remained brightest in the east until at least 9568 days. 
\citet{Zanardo2013} also measured larger radio expansion velocities in the east than the west. 

This behavior suggests the evolution of the remnant has been delayed in the west compared to the east, implying either an asymmetric expansion of the blast wave, asymmetries in the CSM density profile, or both.  The first case implies an asymmetric explosion, while the second case implies asymmetries in the progenitor winds. Interestingly, \citet{Zanardo2014} found residual emission offset slightly west of the ER center in ATCA and ALMA observations which is suggestive of a pulsar wind nebula (PWN); if this tentative PWN emission is confirmed, it would imply the central pulsar received a westward kick along with a corresponding higher energy outflow to the east.

We do not yet detect any emission from a central object. To obtain a simple estimate of the upper limit on the flux, we added a central point source to the best-fit model image and increased the flux until the $\chi^2$ value increased by 2.706, corresponding to the $90\%$ confidence limit.  For the 2015 September observation (Obs ID 16756), we find a limit of $9\times10^{-4}$counts s$^{-1}$ in the 2-10 keV band.  Stacking observations resulted in a maximum 2-10 keV count rate of $6\times10^{-4}$counts s$^{-1}$.  Assuming a nonthermal spectrum with a typical power law of index $\Gamma=1.5$ and $N_H=0.235\times10^{22}$ \nhunit, these translate to $L_{X,16756}\lesssim1.5\times10^{34}$erg s$^{-1}$ and $L_{X,stacked}\lesssim3.1\times10^{33}$erg s$^{-1}$, respectively. \citet{Orlando2015} estimate the local absorbing column density in the center of the ER to be $\sim$$5\times10^{22}$\nhunit, 20 times higher than interstellar absorption along the line of sight and far too high to allow detection of faint emission from a central point source.  \edit1{Assuming this higher absorption, we obtain limits on the intrinsic luminosity of $L_{X,16756}\lesssim3.3\times10^{34}$erg s$^{-1}$ and $L_{X,stacked}\lesssim1.2\times10^{34}$erg s$^{-1}$. In general, the strong X-ray emission from the ER hampers detection of a faint central source; the putative central object can probably not be detected in the X-ray band unless the emission from the ring and other sources (such as shocked ejecta) fades, the internal absorption decreases, or both.}

\section{Conclusions}
\label{section:conclusions}

We report our imaging and photometric results from 31 epochs of \chandra\ observations of SN 1987A, covering 16 years.  Our results are consistent with the overall physical picture of a smooth ER ring with dense clumps embedded around the inner edge. Changes in the soft X-ray light curve and reversal of the east-west asymmetry between 7000 and 8000 days are consistent with the optical and infrared results which demonstrate the end of shock interaction with the dense clumps at this time.  After day 8000 shocks continue to move through the smooth ring component, which has a lower density than the clumps, resulting in increasing X-ray flux until $\sim$9500 days.  The 0.5-2 keV light curve then levels off at $\sim$$8\times10^{-12}$ ergs s$^{-1}$cm${^{-2}}$, while the latest image indicates the eastern side of the ER is beginning to fade, evidence that the blast wave has moved into a lower density region beyond the ER.  Evolution of the morphology implies an asymmetric evolution of the newborn remnant, with the above development delayed in the west compared to the east.  This asymmetry is evidence of asymmetry in the explosion, the CSM density profile, or both. Similarities in the hard X-ray and radio light curves, expansion, and morphologies suggest both the hard X-ray and radio emission originate from the same region, likely the smooth component of the ring.

Future observations of the X-ray light curve and morphology will trace the density profile of the material outside the ER, which is currently unknown and records the history of the progenitor's stellar wind.   These observations can aid in distinguishing between different models of the progenitor's evolution.  Additionally, the impending brightening of the ejecta will soon allow measurements of its composition and structure via \chandra\ observations, placing constraints on properties of the supernova and the progenitor star.  Emission from the reverse-shocked ejecta may also help reveal the origin of the observed east-west asymmetry.


\acknowledgments
The authors would like to thank G. Zanardo for providing the 44 GHz images and P. Broos for assistance with the ACIS pileup correction. The scientific results reported in this article are based on observations made by the Chandra X-ray Observatory and have made use of software provided by the Chandra X-ray Center in the application package CIAO.
Support for this work was provided by the National Aeronautics and Space Administration through Chandra Award Numbers GO3-14058X, GO4-15056X, and GO5-16054X issued by the Chandra X-ray Observatory Center, which is operated by the Smithsonian Astrophysical Observatory for and on behalf of the National Aeronautics Space Administration under contract NAS8-03060.

\bibliography{/Users/kafrank/Dropbox/Mendeley/library_aastex}

\end{document}